\pdfoutput=1  

\documentclass[aip,pof,reprint,groupedaddress,amsmath,amssymb]{revtex4-1}

\usepackage[utf8]{inputenc}
\usepackage[T1]{fontenc}
\usepackage[english]{babel} 

\usepackage{newtxtext}  
\usepackage[varvw]{newtxmath}  
\usepackage{helvet}  

\usepackage{graphicx}  
\usepackage{gensymb}  
\usepackage{bm}  
\usepackage{hyperref}  
\usepackage[all]{hypcap}  

\renewcommand{\vec}[1]{\bm{{#1}}}  

\newcommand{\dimless}{\mathrm}  
\newcommand{\Ra}{\dimless{Ra}}  
\newcommand{\Ha}{\dimless{Ha}}  
\newcommand{\Q}{\dimless{Q}}  
\renewcommand{\Pr}{\dimless{Pr}}  
\newcommand{\Nu}{\dimless{Nu}}  
\renewcommand{\Re}{\dimless{Re}}  
\newcommand{\Rm}{\dimless{Rm}}  
\newcommand{\Pm}{\dimless{Pm}}  

\renewcommand{\citet}[1]{%
  {\protect\NoHyper\citeauthor{#1}\protect\endNoHyper}\citep{#1}}

\begin{document}

\onecolumngrid
\begin{center}\footnotesize
This article may be downloaded for personal use only.
Any other use requires prior permission of the author and AIP Publishing. 

This article appeared in Phys.\ Fluids \textbf{32}, 107101 (2020) and may be found at \url{https://doi.org/10.1063/5.0021895}.
\end{center}
\twocolumngrid

\title{Refined mean field model of heat and momentum transfer in magnetoconvection}

\date{September 9, 2020}

\author{Till Z\"urner}
\email{till.zuerner@ensta-paris.fr}
\homepage[\newline%
          {\includegraphics[height=7pt]{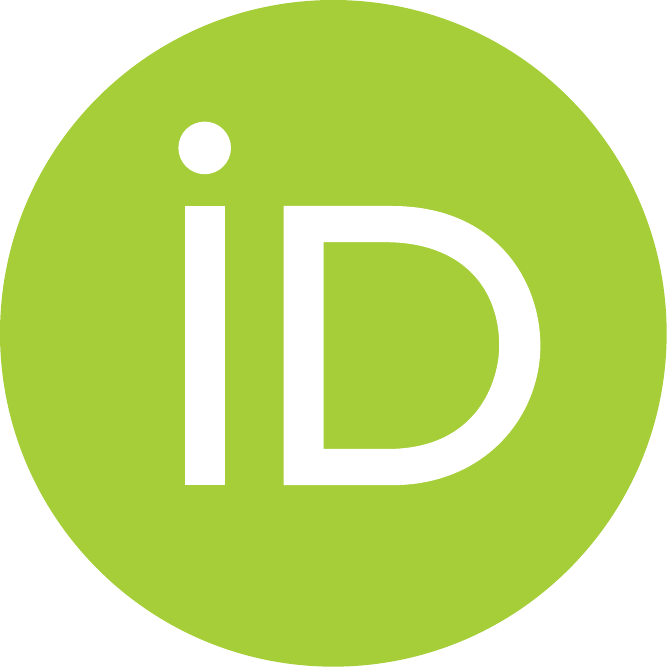}\hspace{2pt}}]
         {https://orcid.org/0000-0001-6488-6611}
\affiliation{%
  Institut des Sciences de la M\'ecanique et Applications Industrielles (IMSIA), \\%
  ENSTA-ParisTech/CNRS/CEA/EDF/Institut Polytechnique de Paris, \\%
  828 Boulevard des Mar\'echaux, 91120 Palaiseau, France}

\begin{abstract}
In this article, the theoretical model on heat and momentum transfer for Rayleigh-B\'enard convection in a vertical magnetic field by Z\"urner \emph{et~al.}\ (Phys.\ Rev.~E \textbf{94}, 043108 (2016)) is revisited.
Using new data from recent experimental and numerical studies the model is simplified and extended to the full range of Hartmann numbers, reproducing the results of the Grossmann-Lohse theory in the limit of vanishing magnetic fields.
The revised model is compared to experimental results in liquid metal magnetoconvection and shows that the heat transport is described satisfactorily.
The momentum transport, represented by the Reynolds number, agrees less well which reveals some shortcomings in the theoretical treatment of magnetoconvection.
\end{abstract}

\maketitle

\section{Introduction}
\label{sec:intro}

Magnetoconvection considers the interaction of magnetic fields with thermal convection flows in electrically conducting fluids.
The most notable examples of such systems in nature are liquid iron cores of planets and the plasma inside stars generating global magnetic fields in the so-called dynamo effect.\citep{Davidson2001,Moffatt2019}
In technological applications, magnetoconvection may be relevant for liquid metal batteries\citep{Kelley2018} and in proposed liquid metal cooling blankets for fusion reactors.\citep{Ihli2008}
The study of magnetoconvection is numerically and experimentally difficult due to the extreme conditions that often govern these systems.
Additionally, the most relevant fluids are liquid metals and plasmas which are either very hard or impossible to handle experimentally.
A theoretical understanding of canonical setups is thus important to understand the relevant mechanisms at play and to predict their behavior beyond the currently accessible parameter space.

In a previous article,\citep{Zurner2016c} a theoretical model was developed to predict the heat and momentum transfer in a Rayleigh-B\'enard convection~(RBC) system subject to a vertical magnetic field.
It utilized the ansatz by \citet{Grossmann2000} and incorporated the effect of Joule dissipation induced by the magnetic field.
The preceding works of \citet{Chakraborty2008} on the same topic should be mentioned here as well.
At the time, the study suffered the lack of numerical and especially experimental data which limited a proper evaluation and validation of the theory.
However, after a number of new studies have been published on the topic over the past few years the model can be revisited and revised.
The aim of the present article is (i)~to simplify the existing model by reducing its number of free parameters as well as by reconsidering the validity of the included physical mechanisms and (ii)~to extend it to a larger parameter space.

\begin{table*}
\centering
\caption{%
  Parameters of experimental and numerical data on RBC with a vertical magnetic field in chronological order.
  Listed are the Prandtl number~$\Pr$ and the range of Rayleigh and Hartmann numbers ($\Ra_\mathrm{min/max}$ and $\Ha_\mathrm{min/max}$).
  Experiments are marked by E and direct numerical simulations by S.
  In addition, the cell aspect ratio~$\Gamma$ is given as $\mathit{diameter}:\mathit{height}$ for cylindrical cells and as $\mathit{width}:\mathit{depth}:\mathit{height}$ for rectangular cells.
  For \citet{Cioni2000}, corresponding data at $\Ha=0$ were published in Ref.~\onlinecite{Cioni1997}.}
\label{tab:data}
\begin{ruledtabular}
\begin{tabular}{clrrrrrc}
  & Reference & 
  \multicolumn{1}{l}{$\Pr$} & 
  \multicolumn{1}{l}{$\Ra_\mathrm{min}$} &
  \multicolumn{1}{l}{$\Ra_\mathrm{max}$} &
  \multicolumn{1}{l}{$\Ha_\mathrm{min}$} & 
  \multicolumn{1}{l}{$\Ha_\mathrm{max}$} & 
  \multicolumn{1}{c}{$\Gamma$} \\[3pt]
  \hline\rule{0pt}{\normalbaselineskip}%
  E & \citet{Cioni2000} & 
  $0.025$ & $2\times 10^7$ & $3\times 10^9$ & $850$ & $1980$ &  $1 : 1$ \\
  E & \citet{Aurnou2001} & 
  $0.025$ & $4\times 10^2$ & $7\times 10^4$ & $26$ & $35$ & $8.3 : 8.3 : 1$ \\
  E & \citet{Burr2001} & 
  $0.020$ & $3\times 10^3$ & $1\times 10^5$ & $10$ & $120$ & $10 : 20 : 1$ \\
  E & \citet{King2015} & 
  $0.024$ & $2\times 10^6$ & $2\times 10^8$ & $0$ & $1110$ & $1 : 1$ \\
  S & \citet{Liu2018} & 
  $0.025$ & $1\times10^7$ & $1\times10^7$ & $0$ & $2000$ & $4 : 4 : 1$ \\
  S & \citet{Yan2019} & 
  $1$ & $1\times10^4$ & $8\times10^{10}$ & $0$ & $10\,000$ & periodic \\
   & & 
  $0.025$ & $2\times10^7$ & $1.7\times10^8$ & $1414$ & $1414$ & periodic \\
  S & \citet{Lim2019} & 
  $8$ & $5\times10^5$ & $1\times10^{10}$ & $0$ & $800$ & $1 : 1 : 1$ \\
  E & \citet{Zurner2020} & 
  $0.029$ & $1\times10^6$ & $6\times10^7$ & $0$ & $1050$ & $1 : 1$ \\
  S & \citet{Akhmedagaev2020} & 
  $0.025$ & $1\times10^7$ & $1\times10^9$ & $0$ & $1400$ & $1 : 1$
\end{tabular}
\end{ruledtabular}
\end{table*}

Rayleigh-B\'enard convection considers a horizontal fluid layer of height~$H$ heated at its lower boundary and cooled at its upper boundary with constant temperatures~$T_\mathrm{bot}$ and~$T_\mathrm{top}$, respectively, where $T_\mathrm{top} < T_\mathrm{bot}$.
A fluid with a sufficiently large electrical conductivity~$\sigma$ can be influenced by imposing a magnetic field which in the present case is a homogeneous vertical magnetic field~$\vec B_0 = B_0 \vec e_z$ (with $z$ as vertical axis).
The flow is controlled by five dimensionless parameters
\begin{equation}
\begin{gathered}
\Ra = \frac{g\alpha\Delta T H^3}{\nu\kappa} \,, \qquad
\Ha = B_0 H \sqrt{\frac{\sigma}{\rho_0\nu}} \,, \\
\Pr = \frac{\nu}{\kappa} \,, \qquad
\Pm = \frac{\nu}{\eta} \,, \qquad
\Gamma = \frac{L}{H} \,.
\end{gathered}
\end{equation}
The Rayleigh number~$\Ra$ quantifies the thermal driving of the fluid by the temperature difference~$\Delta T = T_\mathrm{bot} - T_\mathrm{top}$ and the Hartmann number~$\Ha$ gives a measure of the magnetic field strength.
The fluid is characterized by the thermal Prandtl number~$\Pr$ and the magnetic Prandtl number~$\Pm$ which compare the kinematic viscosity~$\nu$ to the thermal diffusivity~$\kappa$ and the magnetic diffusivity~$\eta = 1/(\mu\sigma)$, respectively.
Lastly, the aspect ratio $\Gamma$ is the ratio of horizontal extent~$L$ of the fluid layer and the layer height~$H$.
The remaining quantities are the acceleration due to gravity~$g$, the magnetic permeability~$\mu$, the volumetric thermal expansion coefficient~$\alpha$ and the mass density~$\rho_0$ of the fluid at a reference temperature~$T_0$.
An alternative parameter to the Hartmann number is the Chandrasekhar number~$\Q = \Ha^2$.
Of major interest in the convection research are the globally averaged quantities of heat and momentum transport, represented by the Nusselt and Reynolds number
\begin{align}
\Nu &= 1 + \frac{H \langle u_z T\rangle}{\kappa\Delta T} \,, &
\Re &= \frac{U H}{\nu} \,,
\end{align}
respectively.
The symbol $\langle\cdot\rangle$ denotes an average over the fluid volume and time.
The characteristic velocity~$U$ is the speed of the mean wind in the convective flow.
It is generally estimated by the root-mean-square~(rms) average of the velocity field~$\vec v = v_i\vec e_i$ over the whole fluid volume $U = \langle v_i^2\rangle^{1/2}$.
Another parameter of the magnetoconvection system is the magnetic Reynolds number~$\Rm = \Pm\Re$.
It compares the advection of the magnetic field by the flow to its diffusion.
More detailed, at high $\Rm > 1$ the magnetic field can be deformed by the flow, while at low $\Rm \ll 1$ alterations to the external field $\vec B_0$ can generally be neglected.\citep{Davidson2001}

Experimental investigations of magnetoconvection require a working fluid with a sufficiently large electrical conductivity.
In the vast majority of cases, liquid metals are the only option fitting this criterion.
Their high electrical conductivity~$\sigma\sim10^6$\,S/m also gives them a good thermal conductivity which places them in the low Prandtl number regime $\Pr \ll 1$.
Experiments with a watery sulfuric acid ($\Pr=12$, $\sigma\sim10^2$\,S/m) do exist,\citep{Aujogue2016} though to reach the same $\Ha$ as in liquid metals magnetic fields of two orders of magnitude higher strength are required.
Flow measurements in liquid metals are very difficult due to their opaque nature and high heat fluxes are necessary to reach large Rayleigh numbers compared to other common fluids such as air or water. 
Notable early works in liquid metal RBC without magnetic field include Ref.~\onlinecite{Takeshita1996,Cioni1997,Glazier1999,Tsuji2005}.
In recent years the topic experienced a number of new experimental efforts.\citep{King2015,Khalilov2018,Vogt2018a,Akashi2019,Zurner2019}
Experiments of RBC including the effects of a vertical magnetic field are much more rare. 
When the initial theory\citep{Zurner2016c} on heat and momentum transport in magnetoconvection was published, only data by \citet{Cioni2000} at high $\Ha \ge 850$ and $\Ra$ up to $3\times10^9$ were available.
Other studies\citep{Aurnou2001,Burr2001} were at very low $\Ra \le 10^5$ and $\Ha \le 120$.
Since then, experimental heat transport data by \citet{King2015} and \citet{Zurner2020} were published, the latter including the currently sole measurements of the velocity field in liquid metal RBC with a vertical magnetic field.
The parameter ranges covered by the now available experimental data are summarized in Table~\ref{tab:data}.

Numerical simulations of turbulent RBC with a vertical magnetic field at low Prandtl numbers are published by \citet{Liu2018}, \citet{Yan2019} and \citet{Akhmedagaev2020} (all at $\Pr = 0.025$).
Simulations at higher $\Pr$ exist by \citet{Yan2019} ($\Pr = 1$) and \citet{Lim2019} ($\Pr = 8$).
Their advantage over experiments is, of course, the full knowledge of the convective velocity field.
However, for small $\Pr$ exhaustive parameter surveys are prohibitively expensive in terms of computation power.
Nonetheless, their detailed insights on magnetoconvection are instrumental in revising the theoretical model.
The parameters of the above publications are listed in Table~\ref{tab:data}.
Some studies focusing on magnetoconvection close to the onset should be mentioned here as well\citep{Basak2015,Rameshwar2017,Mondal2018}.

This article is structured as follows.
The next section~\ref{sec:first_model} recapitulates the central ideas of the Grossmann-Lohse~(GL) approach as the basis of the theoretical model.
In section~\ref{sec:revision} the different aspects of the magnetoconvection model are reviewed.
Where necessary, they are altered or extended.
The updated model is evaluated with the available experimental data and its results are discussed in section~\ref{sec:result}.
Lastly, section~\ref{sec:conclusion} gives the final conclusions and a short discussion.

\section{The magnetohydrodynamic extension of the Grossmann-Lohse model}
\label{sec:first_model}

For completeness, the framework of the theoretical model of heat and momentum transfer in magnetoconvection is outlined here. 
It is based on Ref.~\onlinecite{Zurner2016c} which builds on the original works by Grossmann and Lohse\citep{Grossmann2000,Grossmann2001,Grossmann2002,Grossmann2004,Grossmann2008}, an updated parameter fit by \citet{Stevens2013} (both for the nonmagnetic convection case; see also \citet{Bhattacharya2018} and \citet{Bhattacharya2020} for a slightly modified approach) and investigations by \citet{Chakraborty2008} for the magnetoconvection case.
The GL theory considers the volume- and time-averaged viscous and thermal energy dissipation rates~(DR) -- $\varepsilon_\nu$ and $\varepsilon_\kappa$, respectively -- in the convective flow
\begin{subequations}
\label{eq:DR}
\begin{align}
\label{eq:DR_visc}
\varepsilon_\nu &= \frac{\nu}{2} \left\langle\left(
  \partial_i v_j + \partial_j v_i\right)^2 \right\rangle  \,, \\
\label{eq:DR_therm}
\varepsilon_\kappa
  &= \kappa \left\langle\left(\partial_i T\right)^2 \right\rangle  \,, \\
\label{eq:DR_magn}
\varepsilon_\eta &= \frac{\eta}{2} \left\langle\left(
  \partial_i b_j - \partial_j b_i\right)^2 \right\rangle \,.
\end{align}
\end{subequations}
Here, the Einstein summation convention is used over the coordinates $i,j=x,y,z$ and $\partial_i \equiv \partial/\partial x_i$ is a short notation for the spatial partial derivatives.
In the case of magnetoconvection, the additional magnetic DR~$\varepsilon_\eta$ due to Joule dissipation has to be considered.
Since the imposed magnetic field~$\vec B_0$ is homogeneous, only the secondary magnetic field~$\vec b = b_i\vec e_i$ induced by the interaction of $\vec v$ and $\vec B_0$ is relevant for the calculation of the magnetic DR.
It should be mentioned that the above definition of~$\varepsilon_\eta$ differs by a factor of $1/(\mu\rho_0)$ from other studies.\citep{Chakraborty2008,Zurner2016c}
This is done to have consistent units for the three dissipation rates: $[\varepsilon_\nu] = (\text{m}/\text{s})^2 / \text{s}$, $[\varepsilon_\kappa] = \text{K}^2 / \text{s}$ and $[\varepsilon_\eta] = \text{T}^2 / \text{s}$ with the above definitions.

The GL approach is a mean field theory since only average quantities are considered.
As a result, the aspect ratio $\Gamma$ or the cell geometry is not incorporated explicitly into the theory and the effect of side walls, which can constrain the transport, is neglected.
Only the top and bottom boundaries of the fluid layer are relevant.
They are always assumed to be rigid and electrically insulating which results in a no-slip boundary condition for the velocity field.

The averaged DR in~\eqref{eq:DR} are of importance since in statistically stationary turbulence the exact equations
\begin{align}
\label{eq:base_relations}
\varepsilon_\nu + \frac{\varepsilon_\eta}{\mu\rho_0} &=  
  \frac{\nu^3}{H^4} \frac{(\Nu-1) \Ra}{\Pr^2} \,, &
\varepsilon_\kappa &= \kappa \frac{(\Delta T)^2}{H^2} \Nu 
\end{align}
can be obtained.\citep{Shraiman1990,Chakraborty2008}
In the GL theory, the second term on the left-hand-side of the first equation is not present,\citep{Grossmann2000} since in that case $\varepsilon_\eta = 0$.
The GL approach now splits the DR into their contributions from characteristic regions of the flow, namely the bulk and the boundary layer (BL)
\begin{subequations}
\label{eq:DR_split}
\begin{align}
\label{eq:DR_split_visc}
\varepsilon_\nu
  &= \varepsilon_{\nu,\mathrm{Bulk}} + \varepsilon_{\nu,\mathrm{BL}} \,, \\
\label{eq:DR_split_magn}
\varepsilon_\eta
  &= \varepsilon_{\eta,\mathrm{Bulk}} + \varepsilon_{\eta,\mathrm{BL}} \,, \\
\label{eq:DR_split_therm}
\varepsilon_\kappa
  &= \kappa\frac{(\Delta T)^2}{H^2} + \varepsilon_{\kappa,\mathrm{Bulk}} + 
     \varepsilon_{\kappa,\mathrm{BL}} \,.
\end{align}
\end{subequations}
The term $\kappa(\Delta T)^2/H^2$ in~\eqref{eq:DR_split_therm} is the contribution of pure heat conduction to $\varepsilon_\kappa$.
For high Nusselt numbers this term is often neglected in comparison to the advection based contributions of the bulk and BL regions but becomes relevant in low-$\Nu$ regimes.\citep{Grossmann2008}
Now, the individual contributions in~\eqref{eq:DR_split} are estimated by considering that the bulk dissipation is dominated by inertia and the BL dissipation by viscous effects.
These estimates are then multiplied by free model parameters and combined with~\eqref{eq:base_relations} and~\eqref{eq:DR_split} to form the model equations.
In the present article, the model fit parameters of the GL theory (i.e., $\Ha = 0$) are referred to by capital letters~$A$ and~$C_1$ to~$C_4$ (corresponding to~$a$ and~$c_1$ to~$c_4$ in Ref.~\onlinecite{Stevens2013}) and the parameters of the present magnetoconvection model are denoted by small letters~$a$ and~$c_1$ to~$c_6$.
Note, that the parameters $C_i$ and $c_i$ do not correspond to the same terms.
The initial model\citep{Zurner2016c} utilized the following estimates for the DR contributions
\begin{subequations}
\label{eq:DR_estimate}
\begin{align}
\label{eq:DR_visc_bulk_estimate}
\varepsilon_{\nu,\mathrm{Bulk}} 
  &\sim \frac{U^3}{H}
  = \frac{\nu^3}{H^4} \Re^3 \,, \\
\label{eq:DR_visc_BL_estimate}
\varepsilon_{\nu,\mathrm{BL}} 
  &\sim \nu\frac{U^2}{\delta_{v,B}^2}\,\frac{\delta_{v,B}}{H}
  = \frac{\nu^3}{H^4} \Re^2 \Ha \,, \\
\label{eq:DR_magn_bulk_estimate}
\varepsilon_{\eta,\mathrm{Bulk}} 
  &\sim \eta \frac{\Rm^2 B_0^2}{H^2}
  = \mu\rho_0\frac{\nu^3}{H^4} \Re^2 \Ha^2 \,, \\
\label{eq:DR_magn_BL_estimate}
\varepsilon_{\eta,\mathrm{BL}} 
  &\sim \eta \frac{\Rm^2 B_0^2}{\delta_{v,B}^2} \frac{\delta_{v,B}}{H} 
  = \mu\rho_0\frac{\nu^3}{H^4} \Re^2 \Ha^3 \,, \\
\label{eq:DR_therm_bulk_estimate}
\varepsilon_{\kappa,\mathrm{Bulk}} 
  &\sim \frac{(\Delta T)^2U}{H} 
  = \kappa\frac{(\Delta T)^2}{H^2} \Re \Pr \,, \\
\label{eq:DR_therm_BL_estimate}
\varepsilon_{\kappa,\mathrm{BL}} 
  &\sim \kappa\frac{(\Delta T)^2}{H^2} \sqrt{\Re \Pr} \,.
\end{align}
\end{subequations}
The above scaling relations are based on the following assumptions (more information on their derivation can be found in Appendix~\ref{apx:modelEq_DRest}):
(i)~The Prandtl number is restricted to the $\Pr \ll 1$ case of liquid metals.
(ii)~The Hartmann number is large enough, that the viscous boundary layers at the top and bottom boundary have to be substituted by Hartmann layers.
The viscous BL thickness~$\delta_v$ transforms then to~$\delta_{v,B} = H/\Ha$.
For the $\Ha = 0$ case, the GL theory assumes a Blasius-type BL with a thickness $\delta_{v,0} = aH/\sqrt{\Re}$, where $a$ is a free parameter.\citep{Grossmann2001}
The thermal BL thickness, given by~$\delta_T = H/(2\Nu)$, is unaffected by this assumption.\citep{Akhmedagaev2020}
(iii)~The magnetic Reynolds number is sufficiently low, $\Rm \ll 1$, so that the quasistatic approximation can be applied.
This means that, compared to the external magnetic field~$\vec B_0$, the effect of the induced magnetic field~$\vec b$ on the eddy currents can be neglected.
Since liquid metals have $\Pm\sim10^{-6}$, very high Reynolds numbers of $\Re\sim10^6$ are needed to invalidate this assumption.
In this approximation, the magnitude of $\vec b$ can be estimated as $b \sim \Rm B_0$.\citep{Davidson2001}

Additionally, three regime transitions are introduced to account for changes in the estimates~\eqref{eq:DR_estimate} for different parameter regimes (for more details see Appendix~\ref{apx:modelEq_transition}).
First, the velocity scale within the thermal BL is $U$ if $\delta_T > \delta_v$. 
However, for the case $\delta_T < \delta_v$ the velocity scale $U \delta_T/\delta_v$ has to be used instead.\citep{Grossmann2000}
This change in scaling is introduced\citep{Grossmann2001} by replacing $\Re \to \Re f(\delta_{v,B}/\delta_T)$ in~\eqref{eq:DR_therm_bulk_estimate} and~\eqref{eq:DR_therm_BL_estimate} with the transition function $f(x) = (1+x^4)^{-1/4}$.
Secondly, $\varepsilon_{\nu,\mathrm{Bulk}} \propto \Re^3$ in~\eqref{eq:DR_visc_bulk_estimate} assumes a turbulent flow, while after a transition to a weakly non-linear flow the scaling is better represented by $\varepsilon_{\nu,\mathrm{Bulk}} \propto \Re^2$.
This is facilitated by multiplying $\varepsilon_{\nu,\mathrm{Bulk}}$ by $g(\Re/\Re^\ast)$, where $g(x) = f(1/x)^{-1}$ and $\Re^\ast$ is a model parameter characterizing the position of transition to fully turbulent convection.\citep{Zurner2016c}
The last transition concerns the onset of convection, which is not naturally recovered by the model and is imposed by replacing occurrences of $\Nu-1$ by $(\Nu-1)/h(\Ra/\Ra_\mathrm{c})$ with the transition function $h(x) = 1 - f(x)$.
The critical Rayleigh number~$\Ra_\mathrm{c}$ is calculated in the Chandrasekhar limit\citep{Chandrasekhar1961} as $\Ra_\mathrm{c} = \pi^2\Ha^2$ which is valid for $\Ha \gtrsim 100$.
This last replacement has to be done only in the model equation used to calculate $\Nu$.\citep{Zurner2016c}

\begin{figure*}
\centering
\includegraphics{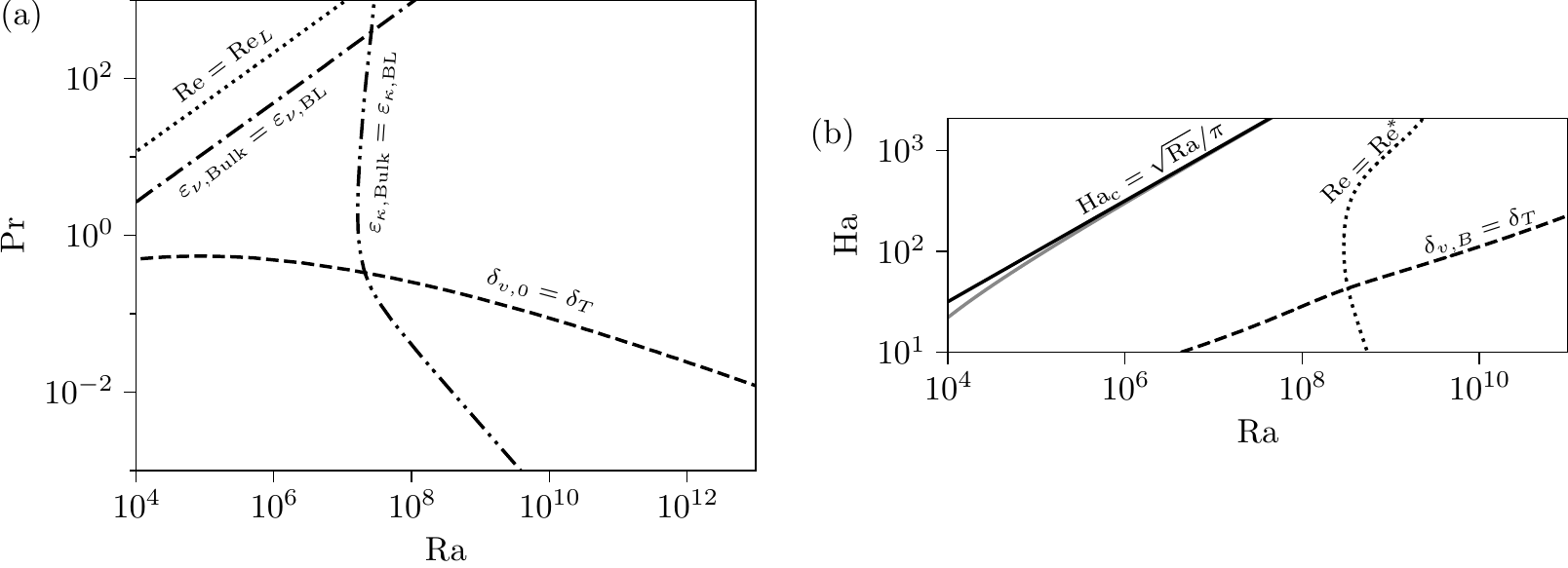}
\caption{%
  (a) Phase diagram of the GL theory at $\Ha=0$ spanned by Rayleigh number $\Ra$ and Prandtl number $\Pr$ according to \citet{Stevens2013}
  Shown are the transition boundaries for the BL crossover $\delta_{v,0} = \delta_T$ (dashed line), the equivalence of bulk and BL dissipation for the viscous DR $\varepsilon_{\nu,\mathrm{Bulk}} = \varepsilon_{\nu,\mathrm{BL}}$ (dash-dotted line) and thermal DR $\varepsilon_{\kappa,\mathrm{Bulk}} = \varepsilon_{\kappa,\mathrm{BL}}$ (dash-double-dotted line) and the transition to the large-$\Pr$ regime at $\Re = \Re_L = 3.4$ (dotted line).
  (b)~$(\Ra, \Ha)$ phase diagram of the initial model for magnetoconvection at $\Pr=0.025$ according to \citet{Zurner2016c}
  Shown are the transition boundaries for the BL crossover $\delta_{v,B} = \delta_T$ (dashed line), the transition to the fully turbulent regime at $\Re = \Re^\ast = 5.6\times10^4$ (dotted line) and the Chandrasekhar limit $\Ha_\mathrm{c} = \sqrt{\Ra}/\pi$ (solid line).
  For comparison, the real solution of $\Ha_\mathrm{c}$ from a linear stability analysis\citep{Chandrasekhar1961} is plotted as a solid gray line.}
\label{fig:phasediagram_GL13+mGL16}
\end{figure*}

With these considerations implemented, the final model equations are calculated by multiplying the estimates~\eqref{eq:DR_visc_bulk_estimate} to~\eqref{eq:DR_therm_BL_estimate} with the free model parameters $c_1$ to $c_6$, respectively, and combining them with~\eqref{eq:base_relations} and~\eqref{eq:DR_split}.
The result is~\citep{Zurner2016c}
\begin{subequations}
\label{eq:model_initial}
\begin{gather}
\label{eq:model_initial_Re}
\Re = \frac{\left(\sqrt{c_6^2 + 4c_5(\Nu-1)} - c_6\right)^2}
           {4c_5^2 \Pr\, f\!\left(2\Nu/\Ha\right)} \,, \\
\label{eq:model_initial_Nu}
\frac{(\Nu-1) \Ra}{\mathcal R^2\Pr^{2}h(\Ra/\Ra_\mathrm{c})} 
  = c_1 \mathcal R\, g\!\left(\frac{\mathcal R}{\Re^\ast}\right) + 
    c_2 \Ha + c_3 \Ha^2 + c_4 \Ha^3 \\
\notag
\text{with} \quad
\mathcal R = 
  \frac{\left(\sqrt{c_6^2 + 4c_5(\Nu-1)/h(\Ra/\Ra_\mathrm{c})} - c_6\right)^2}
       {4 c_5^2 \Pr\, f\!\left(2\Nu/\Ha\right)} \,.
\end{gather}
\end{subequations}
Equation~\eqref{eq:model_initial_Nu} contains $\Nu$, $\Ra$, $\Ha$ and $\Pr$ only.
If the values of the model parameters $c_1$ to $c_6$ and $\Re^\ast$ are known, it can be used to numerically calculate $\Nu$ for a point in the ($\Ra, \Ha, \Pr)$ phase space.
Once $\Nu$ is known, $\Re$ can be obtained from~\eqref{eq:model_initial_Re}.

Since the model parameters $c_1$ to $c_6$ and $\Re^\ast$ are \emph{a priori} unknown, they have to be determined by fitting equations~\eqref{eq:model_initial} to experimental data sets of $(\Ra, \Ha, \Pr, \Nu)$ and at least one data point $(\Ra, \Ha, \Pr, \Nu, \Re)$ including the Reynolds number.
In Ref.~\onlinecite{Zurner2016c}, using the heat transfer data by \citet{Cioni2000} and numerical results for the momentum transport, the parameter values $c_1 = 0.053$, $c_2 = -2.4$, $c_3 = 0.014$, $c_4 = -3.7 \times 10^{-6}$, $c_5 = 0.0038$, $c_6 = 0.47$ and $\Re^\ast = 5.6 \times 10^4$ were obtained.
Figure~\ref{fig:phasediagram_GL13+mGL16} shows the regime diagrams of the GL theory\citep{Stevens2013} at $\Ha = 0$ and of the initial model\cite{Zurner2016c} at $\Pr = 0.025$.
These will be used as reference in the following discussion.

\section{Modifications of the framework}
\label{sec:revision}

The original model can be significantly revised by considering its validity boundaries and which assumptions or mechanisms are applicable in that range of parameters.
Each of the following sections considers one aspect of the initial model equations~\eqref{eq:model_initial}.
Some aspects of the model will be corrected as required and new aspects are introduced.

\subsection{The crossover of the thermal and kinetic BL}
\label{sec:revision_BL}

\begin{figure}
\centering
\includegraphics{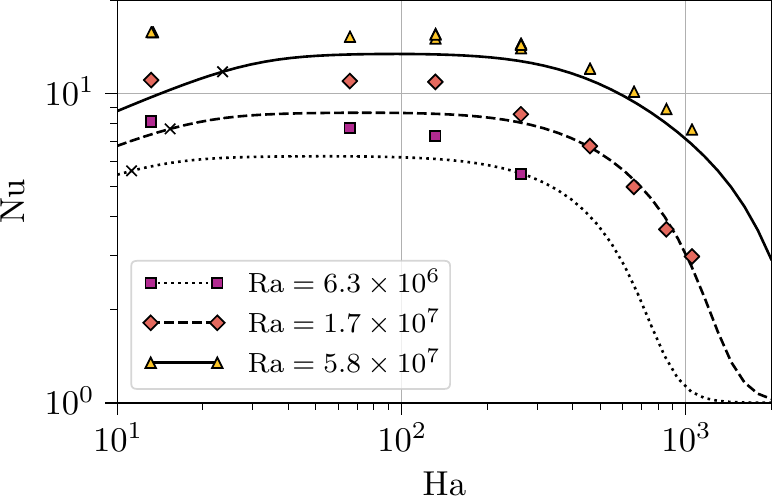}
\caption{%
  Comparison of experimental Nusselt number data\citep{Zurner2020} (filled markers) with the predictions of the initial model\citep{Zurner2016c} (lines) at $\Pr=0.029$ for selected~$\Ra$. 
  The position of the BL crossover $\delta_{v,B} = \delta_T$ is marked by crosses on the respective lines.}
\label{fig:cmp_mGL16_Z20}
\end{figure}

The first topic concerns the velocity scale within the thermal BL.
As discussed in section~\ref{sec:first_model}, the characteristic velocity is chosen as $U$ if $\delta_T > \delta_v$ and as $(\delta_T/\delta_v) U$ if $\delta_T > \delta_v$, which is implemented by the transition function $f(\delta_{v,B}/\delta_T)$ in the initial model~\eqref{eq:model_initial} and by $f(\delta_{v,0}/\delta_T)$ in the GL theory.\citep{Grossmann2001}
This, however, entails an unnecessary complication of the model for low $\Pr$.
Simulations\citep{Scheel2016,Scheel2017} at $\Ha = 0$ and $\Pr=0.025$ show that the viscous BL is smaller than the thermal BL $\delta_v < \delta_T$.
This is also reflected by the results of the GL theory which gives the BL crossover $\delta_{v,0} = \delta_T$ for $\Pr > 0.1$ up to $\Ra = 10^{11}$ (Fig.~\ref{fig:phasediagram_GL13+mGL16}(a)).
By applying a magnetic field, the kinetic BL is decreased due to its eventual transformation\citep{Lim2019} into a Hartmann layer $\delta_{v,B} \propto 1/\Ha$.
Conversely, the thermal boundary layer thickness $\delta_T \propto 1/\Nu$ increases since experiments and simulations in low-$\Pr$ magnetoconvection have shown that $\Nu$ generally decrease for increasing $\Ha$.\citep{Cioni2000,King2015,Liu2018,Zurner2020}
That means that the presence of the $\delta_{v,B} = \delta_T$ regime boundary in the initial model (see dashed line in Fig.~\ref{fig:phasediagram_GL13+mGL16}(b)) is implausible and a result of the insufficient coverage of the low-$\Ha$ regime by the experimental data used for fitting the model parameters.
The discrepancy between the model and experimental data is shown in Fig.~\ref{fig:cmp_mGL16_Z20}.
Measured Nusselt numbers taken from \citet{Zurner2020} (symbols) are compared to the predictions of the initial model~\eqref{eq:model_initial} (lines) at three selected~$\Ra$.
For high $\Ha > 200$, the model captures the experimental results well, but deviates from the experiments at small~$\Ha$.
Especially for $\Ha \to 0$, the experimental $\Nu$~data saturate at a constant value while the model predictions start to decrease.
This coincides with the boundary layer crossover $\delta_{v,B} = \delta_T$ which is marked by a cross on each line.
These considerations show that the BL crossover is not relevant for small $\Pr$ and can actually result in wrong predictions for the low-$\Ha$ regime.
It will thus be eliminated from the revised model equations, i.e., the transition function~$f$ is removed.

The BL crossover only becomes relevant at moderate or high $\Pr$.
Simulations of magnetoconvection\citep{Lim2019} at $\Pr = 8$ found a BL crossover with $\delta_v > \delta_T$ below an optimal Hartmann number.
It is also of interest that the crossover is tied to a short increase of the Nusselt number compared to its value at $\Ha=0$.
As seen in Fig.~\ref{fig:cmp_mGL16_Z20}, the transition function~$f$ emulates such a behavior by generating a local maximum of $\Nu$.
That means, if the present model was to be extended to the intermediate and high $\Pr$ case the transition function~$f$ may be of importance and could be reintroduced.
However, this is not part of the scope of the present work.

\subsection{The limit of small Hartmann numbers}
\label{sec:revision_Ha0}

In the model equations~\eqref{eq:model_initial}, the Hartmann BL~$\delta_{v,B} = H/\Ha$ is used to characterize the kinetic BL. 
This is not applicable for the limit $\Ha\to0$, where the kinetic BL is better described by a Prandtl-Blasius type BL~$\delta_{v,0} = aH/\sqrt{\Re}$ as used by the GL theory.\citep{Grossmann2000}
\citet{Lim2019} proposed a general BL thickness~$\delta_v$ based on a dimensional analysis that connects these two types of BL
\begin{equation}
\label{eq:vBL_transition}
\delta_v 
  = \left(\delta_{v,0}^{-2} + \delta_{v,B}^{-2}\right)^{-1/2}
  = \frac{H}{\sqrt{\Re\, a^{-2} + \Ha^2}} \,.
\end{equation}
For high $\Ha \to \infty$, \eqref{eq:vBL_transition}~becomes a Hartmann layer $\delta_v \to \delta_{v,B}$ and at vanishing magnetic fields the Prandtl-Blasius BL is recovered $\delta_v \to \delta_{v,0}$.
Replacing $\delta_{v,B}$ with $\delta_v$ in the BL contribution of the kinetic and magnetic DR in~\eqref{eq:DR_visc_BL_estimate} and~\eqref{eq:DR_magn_BL_estimate} results in the modified estimates
\begin{subequations}
\label{eq:DR_BL_new-estimate}
\begin{align}
\label{eq:DR_visc_BL_new-estimate}
\varepsilon_{\nu,\mathrm{BL}} 
  &\sim \frac{\nu^3}{H^4} \sqrt{\Re^5 a^{-2} + \Re^4 \Ha^2} \,, \\
\label{eq:DR_magn_BL_new-estimate}
\varepsilon_{\eta,\mathrm{BL}} 
  &\sim \mu\rho_0\frac{\nu^3}{H^4} \sqrt{\Re^5 \Ha^4 a^{-2} + \Re^4 \Ha^6} \,.
\end{align}
\end{subequations}
In the high $\Ha$ limit, these estimates recover the initial scalings~\eqref{eq:DR_estimate}.
For $\Ha\to0$, $\varepsilon_{\eta,\mathrm{BL}}$ vanishes and $\varepsilon_{\nu,\mathrm{BL}}$ becomes the estimate of the GL theory\citep{Grossmann2000} $\varepsilon_{\nu,\mathrm{BL}} \sim (\nu^3/H^4) \Re^{5/2}$.

\subsection{Transition towards laminar bulk flow and onset of convection}
\label{sec:revision_laminar}

The GL ansatz~\citep{Grossmann2000} assumes the existence of a turbulent large-scale wind of velocity~$U$ in the convection cell.
Even with the subsequent extension towards a laminar high-$\Pr$ case\citep{Grossmann2001}, the scaling of the viscous bulk DR has always been assumed to be dominated by inertia ($\varepsilon_\mathrm{\nu, Bulk} \propto \Re^3$).
The initial model for magnetoconvection\citep{Zurner2016c} introduced a transition between the turbulent $\Re^3$-scaling towards a laminar $\Re^2$-scaling of $\varepsilon_\mathrm{\nu, Bulk}$ at a characteristic Reynolds number of $\Re^\ast$ which was evaluated to $\Re^\ast = 5.6 \times 10^4$.
The phase diagram in Fig.~\ref{fig:phasediagram_GL13+mGL16}(b) shows this transition to happen at $\Ra > 10^8$ for all $\Ha$.
However, especially for the $\Ha = 0$ case it is well-known that turbulence in low-$\Pr$ convection sets in at much smaller Rayleigh numbers.\citep{Busse1978,Breuer2004,Schumacher2015}
Since the bulk turbulence is a central assumption of the model, it is evident that this scaling transition on its own is insufficient to model the weakly non-linear and laminar regimes at high~$\Ha$.
The transition function~$g$ and the model parameter $\Re^\ast$ are consequently removed from the model equations.
 
The onset of convection cannot be recovered intrinsically by the current model and would require a proper treatment of the non-turbulent regimes with a complete overhaul of the model ansatz.
This, however, is beyond the scope of this study.
The transition towards the purely conductive regime was previously imposed at the Chandrasekhar limit by a fixed transition function~$h$ in Ref.~\onlinecite{Zurner2016c}.
This approach will be retained and the results of the revised model with and without the imposed onset transition are compared in section~\ref{sec:result}.

The critical Rayleigh number in the Chandrasekhar limit $\Ra_\mathrm{c} = \pi^2\Ha^2$ is valid only for $\Ha \gtrsim 100$.
To allow for the limit $\Ha \to 0$ discussed in the previous section, the argument of the onset transition function~$h$ is replaced based on the critical Hartmann number: $h(\Ha_\mathrm{c}^2 / \Ha^2)$.
In the Chandrasekhar limit, $\Ha_\mathrm{c} = \sqrt{\Ra}/\pi$ which is valid for $\Ra \gtrsim 2\times 10^5$ and all $\Ha$ with a deviation of $\le 10\,\%$ from the proper solution obtained by a linear stability analysis.\citep{Chandrasekhar1961}
Since $\Ha_\mathrm{c}^2/\Ha^2 = \Ra/\Ra_\mathrm{c}$ in the Chandrasekhar limit, this change has only an effect on the validity boundaries of the model.

Recently, simulations\citep{Liu2018} and experiments\citep{Zurner2020} proved the existence of convective flows for $\Ha > \Ha_\mathrm{c}$ concentrated near the lateral walls of the convection cell.
They are denoted as wall modes that cannot be included in the present mean-field theory which neglects the effect of side-walls.

\begin{widetext}
\subsection{Revised model equations}
\label{sec:revision_model}

The initial and new DR contribution estimates~\eqref{eq:DR_estimate} and~\eqref{eq:DR_BL_new-estimate} are multiplied by the model parameters $c_1$ to $c_6$ and combined with equations~\eqref{eq:base_relations} and~\eqref{eq:DR_split}

\begin{gather}
\label{eq:model1}
\frac{(\Nu-1) \Ra}{\Pr^{2}} 
  = c_1 \Re^3
  + c_2 \sqrt{\Re^5 a^{-2} + \Re^4 \Ha^2}
  + c_3 \Re^2 \Ha^2
  + c_4 \sqrt{\Re^5 \Ha^4 a^{-2} + \Re^4 \Ha^6} \,, \\
\label{eq:model2}
\Nu-1 = c_5 \Re \Pr + c_6 \sqrt{\Re \Pr}\,.
\end{gather}
\end{widetext}

For the case $\Ha = 0$, the second model equation~\eqref{eq:model2} does not change while equation~\eqref{eq:model1} becomes
\begin{equation}
\label{eq:model1_Ha0}
\frac{(\Nu-1) \Ra}{\Pr^{2}} 
  = c_1 \Re^3 + \frac{c_2}{a} \Re^{5/2} \,.
\end{equation}
\eqref{eq:model1_Ha0}~and \eqref{eq:model2}~are equal to the GL model equations\citep{Stevens2013} for the low-$\Pr$ regime, i.e., if the regime transitions for the BL crossing and for the high-$\Pr$ limit are removed.
Consequently, the parameters in~\eqref{eq:model1_Ha0} and~\eqref{eq:model2} can be identified with the values of the GL theory\citep{Stevens2013}
\begin{equation}
\label{eq:param_applyGL}
\begin{gathered}
a = A = 0.922 \,, \\
\begin{aligned}
c_1 &= C_2 = 1.38 \,, &
c_2 &= AC_1 = 7.42 \,, \\
c_5 &= C_4 = 0.0252 \,, & \quad
c_6 &= C_3 = 0.487 \,,
\end{aligned}
\end{gathered}
\end{equation}
with $C_1 = 8.05$.
The only remaining unknown parameters are thus $c_3$ and $c_4$.
This is a significant reduction of the number of free parameters compared to the seven fit coefficients of the initial model.
The two remaining parameters need to be fitted to experimental data (see section~\ref{sec:result}).
The original fit\citep{Zurner2016c} resulted in some negative parameter values.
Since only positive coefficients are physically sensible for dissipation rates, the bounds $(0, \infty)$ are imposed on the two parameters during the fitting process.

The initial model for magnetoconvection as well as the GL theory suffer from an ambiguity, where the magnitude of the Reynolds number cannot be determined by heat transport data $(\Ra, \Ha, \Pr, \Nu)$ alone:
$\Re$ can be re-scaled by an arbitrary constant factor without affecting the predicted value of the Nusselt number (see Appendix~\ref{apx:Re_rescale} for more details).
As a result, at least one full data-set $(\Ra, \Ha, \Pr, \Nu, \Re)$ is required to fix the magnitude of $\Re$.
By identifying the the parameters $a$, $c_1$, $c_2$, $c_5$ and $c_6$ with the coefficients of the original GL theory in \eqref{eq:param_applyGL}, this ambiguity has already been resolved for the revised model.
The implications of this choice of parameters on the Reynolds number are discussed in section~\ref{sec:cmp_exp} by comparing it to experimental data.

To fit the model equations to data sets of $(\Ra, \Ha, \Pr, \Nu)$, the Reynolds number is eliminated from \eqref{eq:model1} using \eqref{eq:model2}.
To impose the onset of convection, $\Nu-1$ is replaced\citep{Zurner2016c} by $(\Nu-1) / h(\Ha_\mathrm{c}^2/\Ha^2)$ where $h(x) = 1-(1+x^4)^{-1/4}$
\begin{widetext}
\begin{equation}
\begin{gathered}
\label{eq:model_Nu}
\begin{aligned}
\frac{(\Nu-1) \Ra}{h(\Ha_\mathrm{c}^2/\Ha^2)\Pr^{2}}
  = C_2 \mathcal R^3
    + AC_1 \sqrt{\mathcal R^5 A^{-2} + \mathcal R^4 \Ha^2} 
    + c_3 \mathcal R^2 \Ha^2
    + c_4 \sqrt{\mathcal R^5 \Ha^4 A^{-2} + \mathcal R^4 \Ha^6} \,,
\end{aligned} \\
\text{with} \qquad
\mathcal R = 
  \frac{\left(\sqrt{C_3^2+4C_4(\Nu-1)/h(\Ha_\mathrm{c}^2/\Ha^2)} - C_3\right)^2}
       {4 C_4^2 \Pr} \,.
\end{gathered}
\end{equation}
\end{widetext}
Once the values of $c_3$ and $c_4$ are determined, \eqref{eq:model_Nu}~can be numerically solved for $\Nu$ with a given set of $(\Ra, \Ha, \Pr)$.
The corresponding Reynolds number then follows from~\eqref{eq:model2}
\begin{equation}
\label{eq:model_Re}
\Re = \frac{\left(\sqrt{C_3^2+4C_4(\Nu-1)} - C_3\right)^2}{4 C_4^2 \Pr} \,.
\end{equation}

\section{Results}
\label{sec:result}

The model equation \eqref{eq:model_Nu} is fitted to the experimental data sets $(\Ra,\Ha,\Pr,\Nu)$ by \citet{Cioni2000}, \citet{King2015} and \citet{Zurner2019}
The data by \citet{Aurnou2001} and \citet{Burr2001} are not used, since they have no data $\Ra > 2\times10^5$ which is required by the Chandrasekhar limit approximation (section~\ref{sec:revision_laminar}).
The resulting parameter values of the model including the onset of convection are
\begin{align}
\label{eq:param_fit_onset}
c_3 &= 0.0449 \,, &
c_4 &= 7.52 \times 10^{-18} \approx 0 \,.
\end{align}
The fit returns a standard deviation of 0.0044 for $c_3$, i.e., a relative uncertainty of 10\,\%.
This is a large improvement compared to the initial model, in which the parameter had relative uncertainties of $\sim100\,\%$ due to the lack of data.\citep{Zurner2016c}
The $c_4$ parameter has a fit standard deviation of $2.74\times10^{-6}$.
Compared with its nominal value in~\eqref{eq:param_fit_onset}, $c_4$ can thus be treated as zero, i.e., $\varepsilon_{\eta,\mathrm{BL}}$ has no influence on the result.
This means either that the effect of Joule dissipation in the viscous BL is negligible or that it is only relevant at high magnetic fields $\Ha > 2000$, beyond the currently available experiments (see Table~\ref{tab:data}).
The latter could be the case, since $\varepsilon_{\eta,\mathrm{BL}} \propto \Ha^3$ may increase significantly for high $\Ha$.

If the onset of convection is excluded (i.e., $h(\Ha_\mathrm{c}^2/\Ha^2) \equiv 1$), the fitted parameter values become $c_3 = 0.0520 \pm 0.0058$ and $c_4 = 5.19\times10^{-19} \pm 3.66\times10^{-6} \approx 0$.
The difference between the model predictions with and without imposed onset is investigated in the following section~\ref{sec:cmp_exp} (see also Figs.~\ref{fig:cmp_Nu_onset} and~\ref{fig:cmp_Re}(b)).

Lifting the fitting boundaries of $(0, \infty)$ results in a small negative fit value of $c_4 = (-9.56 \pm 0.26) \times 10^{-5}$ and $c_3 = 0.0599\pm0.0044$.
The negative parameter causes the numerical solution of the model to become unstable for intermediate~$\Ha$.
For example, in the phase space covered by the experimental data the predictions of $\Nu$ and $\Re$ at $\Ha \sim 1000$ can reach magnitudes of 10 times their value at $\Ha = 0$.
The model only produces sensible results in the low- and high-$\Ha$ regime, where the $c_4$-term in~\eqref{eq:model_Nu} has no influence (as the term vanishes for $\Ha \to 0$ and the model is dominated by the imposed transition function~$h$ for large~$\Ha$).
This reinforces the choice to explicitly restrict~$c_3$ and~$c_4$ to positive values and that $c_4$ vanishes.

\subsection{Comparison with experimental data}
\label{sec:cmp_exp}

\begin{figure*}
\centering
\includegraphics{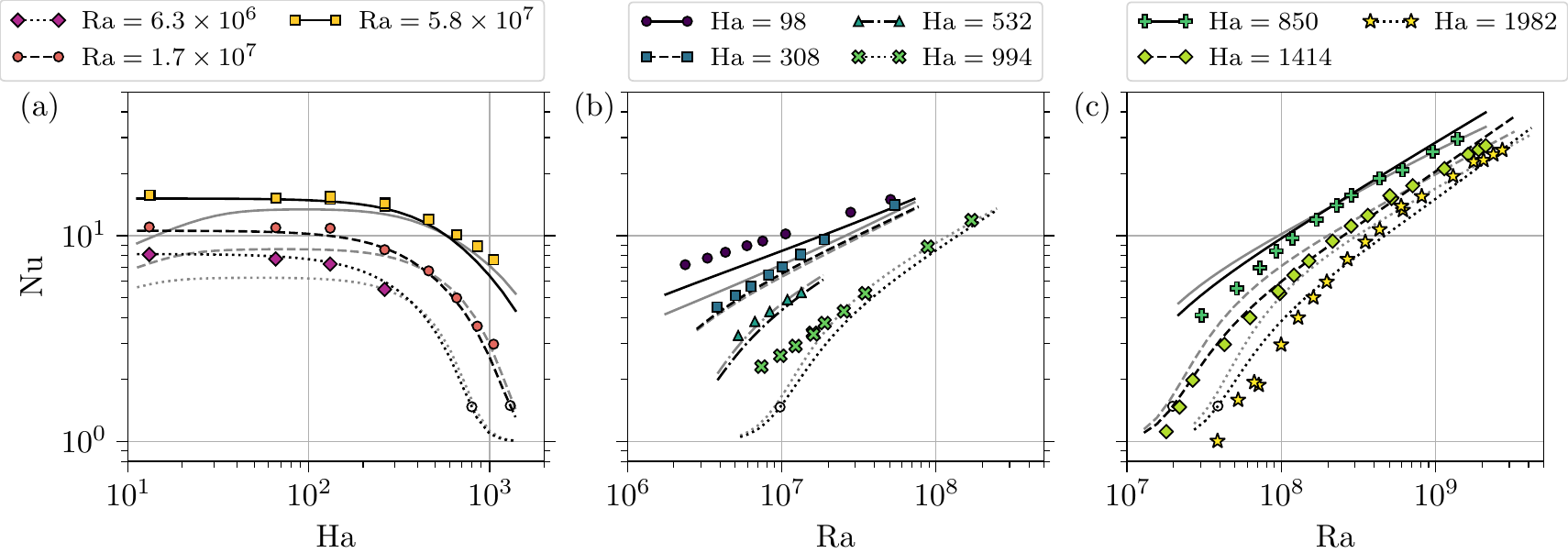}
\caption{%
  Comparison of Nusselt number data from experiments (filled markers) with theoretical predictions (lines).
  Gray lines are the result of the initial model\citep{Zurner2016c} and black lines are the revised model with imposed onset of convection.
  The position of the Chandrasekhar limit is indicated by an open circle on the respective black line.
  The experimental data are selected from the data sets used for the fitting of the model: (a)~\citet{Zurner2020} ($\Nu$ vs.\ $\Ha$ at selected $\Ra$ and $\Pr=0.029$), (b)~\citet{King2015} ($\Nu$ vs.\ $\Ra$ at selected $\Ha$ and $\Pr=0.024$), (c)~\citet{Cioni2000} ($\Nu$ vs.\ $\Ra$ at selected $\Ha$ and $\Pr=0.025$).
  The gray lines and markers in (a) are identical to the lines and markers in Fig.~\ref{fig:cmp_mGL16_Z20}.}
\label{fig:cmp_Nu}
\end{figure*}

\begin{figure*}
\centering
\includegraphics{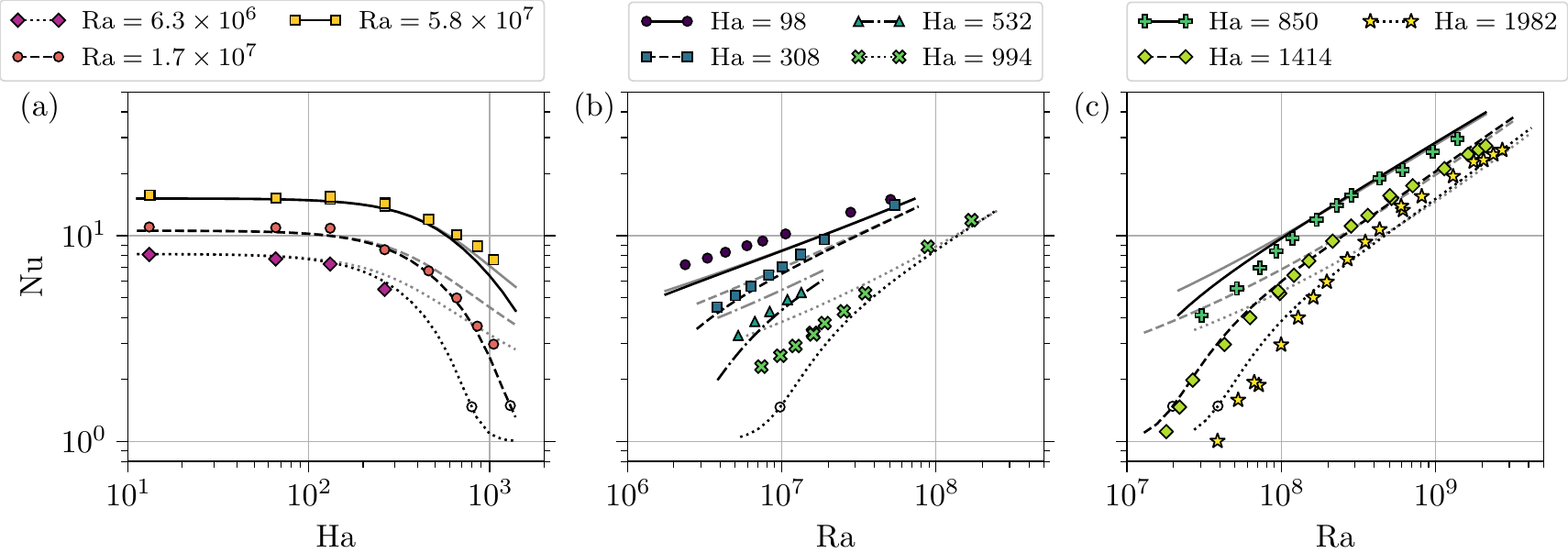}
\caption{%
  Comparison of Nusselt number predictions of the revised model with (black lines) and without (gray lines) imposed onset of convection.
  The position of the Chandrasekhar limit is indicated by an open circle on the respective black line.
  The experimental data (filled markers) and black lines are identical to Fig.~\ref{fig:cmp_Nu}.
  (a)~$\Nu$ vs.\ $\Ha$ at selected $\Ra$ and $\Pr=0.029$ with measurements by \citet{Zurner2020}
  (b)~$\Nu$ vs.\ $\Ra$ at selected $\Ha$ and $\Pr=0.024$ with measurements by \citet{King2015}
  (c)~$\Nu$ vs.\ $\Ra$ at selected $\Ha$ and $\Pr=0.025$ with measurements by \citeauthor{Cioni2000}.\citep{Cioni2000}}
\label{fig:cmp_Nu_onset}
\end{figure*}

\begin{figure*}
\centering
\includegraphics{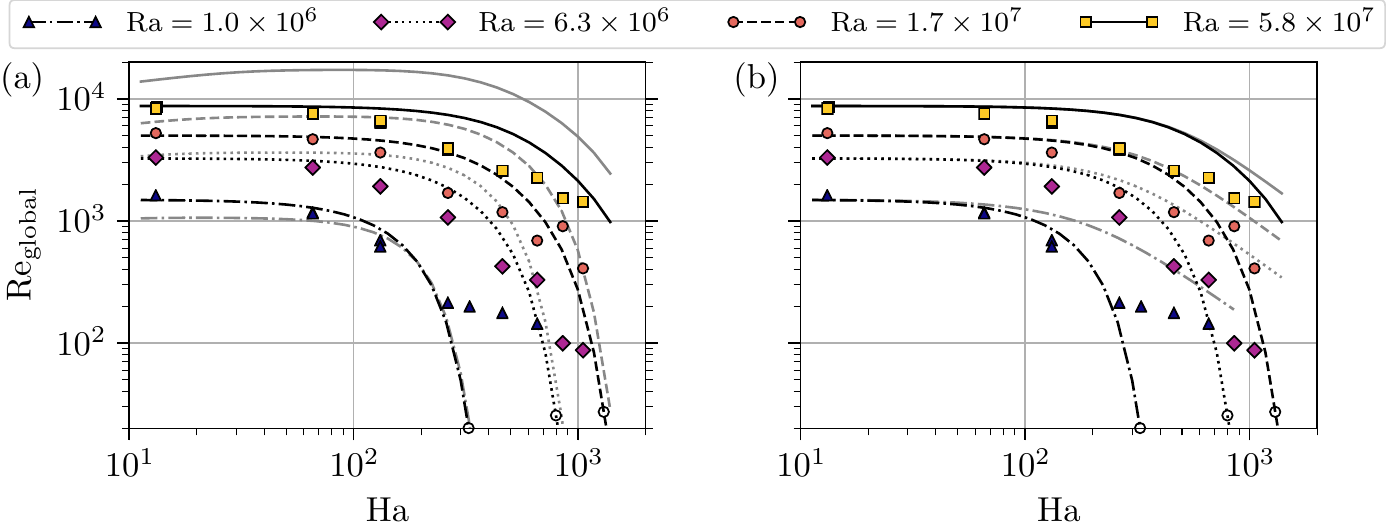}
\caption{%
  Comparison of Reynolds number data from experiments (filled markers) with theoretical predictions (lines).
  Shown is $\Re$ vs.\ $\Ha$ at selected $\Ra$ and $\Pr=0.029$ with measurements by \citet{Zurner2020} ($\Re_\mathrm{global}$ is based on a rms-average over all velocity data measured in the experiment).
  The experimental data correspond to the Nusselt number measurements shown in Figs.~\ref{fig:cmp_Nu}(a) and~\ref{fig:cmp_Nu_onset}(a).
  The black lines show the results of the revised model with imposed onset of convection.
  The position of the Chandrasekhar limit is indicated by an open circle on the respective black line.
  The gray lines are the results of (a)~the initial model\citep{Zurner2016c} and (b)~the revised model without the imposed onset of convection.
  Filled markers and black lines are identical in panels~(a) and~(b).}
\label{fig:cmp_Re}
\end{figure*}

Figure~\ref{fig:cmp_Nu} compares the Nusselt number calculated from the model equation~\eqref{eq:model_Nu} with the experimental data (filled markers) used for fitting the parameters~$c_3$ and~$c_4$.
Plotted are the theoretical predictions of the revised model including the onset of convection (black lines) and of the initial model\citep{Zurner2016c} (gray lines).

The data by \citet{Zurner2020} in Fig.~\ref{fig:cmp_Nu}(a) are well reproduced by the revised model.
It correctly reproduces the saturation of $\Nu$ for low $\Ha\to0$ while the initial model declines as discussed previously in Fig.~\ref{fig:cmp_mGL16_Z20}.
Close to the onset of convection both models give similar predictions as the onset transition function~$h$ dominates the results.
In this range of high $\Ha$, the revised model slightly underpredicts the experimental data.

Nusselt number data by \citet{Cioni2000} (Fig.~\ref{fig:cmp_Nu}(c)) also fit well with the theoretical results.
Here, the revised model generally overpredicts the experimental data near the onset of convection.
The exact progression of $\Nu$ with increasing $\Ra$ is not exactly the same, with the revised model approaching a straight line (i.e., a power law) while the experiment is showing a curvature, but the general trend and magnitude are recovered.
The initial model reproduces the slope of the experimental data better for the highest $\Ra$.
This is not surprising since these were the only data used to fit the parameter of the initial model in Ref.~\onlinecite{Zurner2016c}.
Then again, this better agreement does not extend to low $\Ra$ near the onset of convection where the initial model has even higher values than the revised model.

The data by \citet{King2015} (Fig.~\ref{fig:cmp_Nu}(b)) agree less well with the revised model which approaches the data with increasing $\Ra$, but consistently underpredicts the experiments.
Especially for $\Ha=994$, the onset of convection at the Chandrasekhar limit (at $\Ra_\mathrm{c} = 9.7\times 10^6$, marked by an open circle) is not visible in the experiment.
This is in stark contrast to the other experiments (e.g.\ the $\Ha = 850$ data in Fig.~\ref{fig:cmp_Nu}(c)).
The initial model produces very similar results except for the lowest $\Ha = 98$, where its predictions are even lower than the revised model.
This deviation for low $\Ha$ is in agreement with the previously discussed data in Fig.~\ref{fig:cmp_Nu}(a).

Figure~\ref{fig:cmp_Nu_onset} replots the experimental data (filled markers) and the revised model with imposed onset (black lines) from Fig.~\ref{fig:cmp_Nu} but compares them to the revised model without the imposed onset of convection (gray lines).
The two revised models are identical for low~$\Ha$ (Fig.~\ref{fig:cmp_Nu_onset}(a)) and show only slight differences for large~$\Ra$ (Fig.~\ref{fig:cmp_Nu_onset}(b) and~(c)).
However, close to the Chandrasekhar limit they deviate strongly from one another.
This shows that the model is not intrinsically applicable outside the regime of turbulent convection and why the Chandrasekhar limit is imposed explicitly using a transition function.

Experimental data for the Reynolds number are available from the experiments by \citet{Zurner2020}
The velocity field is probed using ten ultrasound Doppler velocimetry~(UDV) sensors and a characteristic global velocity scale is determined by calculating the rms-average of the measured velocities over time and over all sensors.
The resulting global Reynolds number~$\Re_\mathrm{global}$ is compared to the theoretical predictions in Fig.~\ref{fig:cmp_Re}.
In the low-$\Ha$ limit, the revised model with onset (black lines) correctly reproduces the saturation of $\Re$ at a constant value.
This saturation value is the same as the measured values of $\Re_\mathrm{global}$, though this should be interpreted as purely coincidental.
As shown in Ref.~\onlinecite{Zurner2019}, the GL theory underpredicts Reynolds numbers based on the turbulent large-scale wind in low-$\Pr$ convection by nearly a factor of two.
Due to the choice~\eqref{eq:param_applyGL} of model parameters, this discrepancy is also present in the revised model for magnetoconvection.
At the same time, the magnitude of $\Re_\mathrm{global}$ is affected by many low-velocity areas of the flow that lie within the measurement volume of the UDV sensors.
Coincidentally, this also reduces its magnitude by a factor of about two compared to a wind-based Reynolds number.\citep{Zurner2020}
Nonetheless, the agreement between $\Re_\mathrm{global}$ and the theoretical predictions for small $\Ha$ shows, that the scaling of the Reynolds number is correctly recovered for $\Ha\to0$ while its magnitude only deviates by a constant factor at these $\Pr$.
It is possible to re-scale the theoretical Reynolds number to match the wind-based Reynolds number from experiments at low $\Pr$.
The effects of such a modification are discussed in Appendix~\ref{apx:Re_rescale}.

By comparing Figs.~\ref{fig:cmp_Nu}(a) and~\ref{fig:cmp_Re}(a), it can be seen that the Reynolds number measurements start to decrease from their value at $\Ha = 0$ at lower $\Ha$ than the Nusselt number\citep{Zurner2020}, i.e., they have a shorter saturation plateau.
Since in the model equation~\eqref{eq:model_Re} $\Re$ is directly linked to $\Nu$, the revised model does not recover this behavior and the predicted $\Re$ stays constant up to the same value of $\Ha$ where $\Nu$ starts to drop off.
Also, the onset transition function forces $\Re$ to drop off very fast when approaching the Chandrasekhar limit while $\Re_\mathrm{global}$ decreases at a much slower rate due to the presence of wall modes past the Chandrasekhar limit.\cite{Liu2018,Zurner2020}
The main issue is that in magnetoconvection the degree of turbulence in the flow is not directly linked to the magnitude of the average velocity magnitude.
A combination of high~$\Ra$ and high~$\Ha$ might produce the same Reynolds number of the large scale flow as another combination of low~$\Ra$ and low~$\Ha$, but with weaker turbulent fluctuations.\citep{Zurner2020}
This disconnects the progression of~$\Re$ from that of~$\Nu$, the latter being dependent on the mean velocity magnitude as well as velocity fluctuations.\citep{Lim2019}
This effect is not included in the model equations and as a result, the model does not recover the progression of the Reynolds number correctly, except for the low-$\Ha$ limit.

However, even with these discrepancies the revised model represents an improvement over the results of the initial model (gray lines in Fig.~\ref{fig:cmp_Re}(a)).
The initial model was fitted using numerical Reynolds number data\citep{Zurner2016c} based on the rms-velocity over the whole fluid volume.
As discussed before, it is expected that the resulting predictions are approximately twice as high as the $\Re_\mathrm{global}$ data.
This is only true for the largest measured $\Ra > 10^7$, whereas for lower $\Ra$ the initial model can even underpredict the measurements.
Additionally, the initial $\Re$ predictions starts to decrease for $\Ha\to0$ instead of saturating at a constant value, albeit not as prominently as for $\Nu$ (Fig.~\ref{fig:cmp_Nu}(a)).
Close to the Chandrasekhar limit, the initial and the revised model converge to the same solution due to the effect of the imposed onset transition function.

The revised model without imposed onset of convection is shown in Fig.~\ref{fig:cmp_Re}(b) as gray lines.
Like for the case of $\Nu$ (Fig.~\ref{fig:cmp_Nu_onset}(a)), $\Re$ drops off much more slowly for increasing $\Ha$ than with an imposed transition function.
One could observe that the slope of the lines past the Chandrasekhar limit are similar to the the decrease of the experimental data, albeit at a higher magnitude.
However, since these data are dominated by wall modes which the model cannot take into account, this similarity should be seen as coincidental.

In conclusion, the revised model represents an improvement over the initial model for nearly the whole range of $\Ra$ and $\Ha$.
Especially the extension into the low-$\Ha$ limit has been implemented successfully for both $\Re$ and $\Nu$.
In light of the discrepancies between the experimental data (e.g.\ compare $\Ha = 994$ and 850 in Figs.~\ref{fig:cmp_Nu}(b) and~(c)), the revised model manages to create a satisfactory reproduction of the Nusselt number for all experimental data sets.
In contrast, the Reynolds number predictions do not agree as well with the experimental data.
This points to some mechanisms in magnetoconvection (e.g.\ the suppression of turbulence) that are not yet properly treated in the model.
Lastly, the necessity to impose the onset of convection through a transition function highlights that the weakly non-linear and laminar regimes need to be addressed separately to intrinsically reproduce the Chandrasekhar limit.

\subsection{Regime diagram and validity boundaries}

\begin{figure*}
\centering
\includegraphics{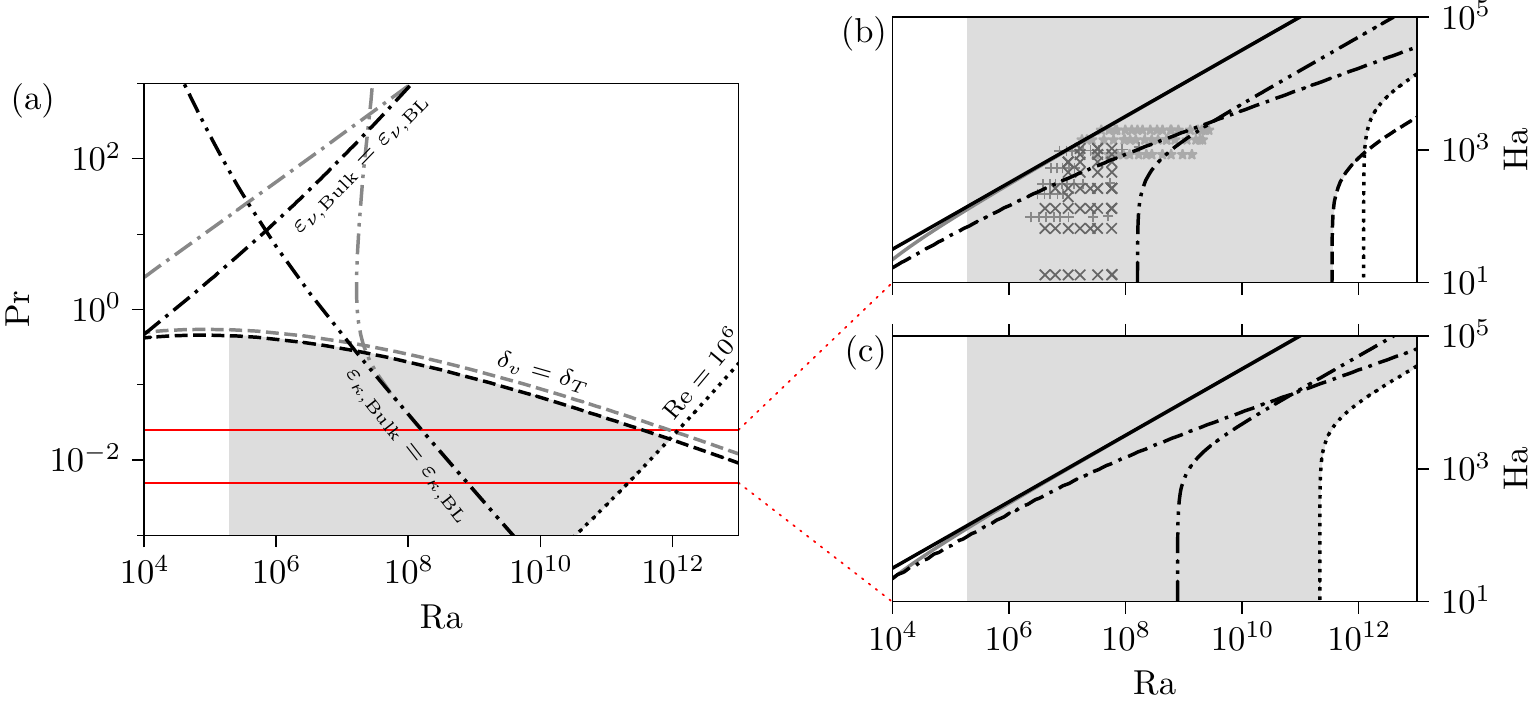}
\caption{%
  Regime diagram and validity boundaries (gray shaded areas) of the revised magnetoconvection model including the onset of convection.
  (a)~$(\Ra, \Pr)$ phase diagram for $\Ha = 0$.
  Shown are the transition boundaries for the BL crossover $\delta_{v} = \delta_T$ (dashed line), $\Re = 10^6$ (dotted line) and the equivalence of bulk and BL dissipation for the viscous DR $\varepsilon_{\nu,\mathrm{Bulk}} = \varepsilon_{\nu,\mathrm{BL}}$ (dash-dotted line) and thermal DR $\varepsilon_{\kappa,\mathrm{Bulk}} = \varepsilon_{\kappa,\mathrm{BL}}$ (dash-double-dotted line).
  The corresponding regime boundaries of the GL theory\citep{Stevens2013} are re-plotted as gray lines from Fig.~\ref{fig:phasediagram_GL13+mGL16}(a). 
  (b)~and (c)~$(\Ra, \Ha)$ phase diagram for $\Pr = 0.025$ (mercury, gallium) and $\Pr = 0.005$ (sodium), respectively.
  These Prandtl numbers are marked in~(a) by horizontal lines.
  The critical Hartmann number~$\Ha_\mathrm{c}$ is displayed in the Chandrasekhar limit (solid black line) and as the rigorous linear stability solution (gray solid line).
  The remaining lines correspond to the regime boundaries in~(a).
  The gray markers in~(b) indicate the experiments used to fit the revised model: \citet{Cioni2000} (stars), \citet{King2015} (pluses) and \citet{Zurner2020} (crosses).
  They mark the region of support for the parameter fit -- all other areas are extrapolated by the model from this region.}
\label{fig:phasediagram_mGL}
\end{figure*}

Figure~\ref{fig:phasediagram_mGL}(a) shows the phase diagram of the revised  model at $\Ha = 0$ (black lines) in comparison to the GL theory (gray lines).
The BL crossover ($\delta_v = \delta_T$, dashed line) is positioned at only slightly smaller $\Pr$ than the result of the GL theory.
Below this line ($\delta_v < \delta_T$), the regime boundaries of the thermal DR contribution crossover ($\varepsilon_{\kappa,\mathrm{Bulk}} = \varepsilon_{\kappa,\mathrm{BL}}$, dash-double-dotted line) coincide for the revised model and the GL theory.
This is expected since the model recovers the GL model equations in the low-$\Pr$ limit.
For $\Pr$ above the BL crossover ($\delta_v > \delta_T$), this regime boundary and the kinetic DR contribution crossover ($\varepsilon_{\nu,\mathrm{Bulk}} = \varepsilon_{\nu,\mathrm{BL}}$, dash-dotted line) deviate strongly from one another, indicating that the model is not applicable for these regimes.

The validity of the model is limited by the following assumptions.
(i)~The working fluid has a low Prandtl number $\Pr \ll 1$. 
This implies that $\delta_v < \delta_T$ and the BL crossover is thus considered as regime boundary for the low-$\Pr$ regime.
(ii)~The Chandrasekhar limit is assumed for the critical Hartmann number~$\Ha_\mathrm{c} = \sqrt{\Ra}/\pi$ which is valid for $\Ra > 2 \times 10^5$.
(iii)~The quasistatic approximation applies, i.e., $\Rm \ll 1$.
For a typical magnetic Prandtl number for liquid metals of $\Pm\sim10^{-6}$, this implies that a Reynolds number of $\Re\sim10^6$ has to be reached to violate this assumption.
The gray shaded area in Fig.~\ref{fig:phasediagram_mGL}(a) shows the validity range of the revised model defined by the above boundaries~(i) to~(iii).
For decreasing and increasing $\Ra$, the limiting boundaries are $\Ra = 2 \times 10^5$ and $\Re = 10^6$ (dotted line), respectively.
With increasing $\Pr$, the model reaches up to the $\delta_v = \delta_T$ boundary (dashed line).

Turning to the case $\Ha > 0$, Figs.~\ref{fig:phasediagram_mGL}(b) and~(c) show the $(\Ra, \Ha)$ phase diagrams for the characteristic Prandtl numbers $\Pr = 0.025$ (mercury, gallium) and $\Pr = 0.005$ (sodium), respectively.
Displayed are the same regime boundaries as in Fig.~\ref{fig:phasediagram_mGL}(a) together with the Chandrasekhar limit (solid black line).
Since the revised model is applicable for all Hartmann numbers, there are no vertical boundaries to the validity range (gray shaded areas).
For low $\Ra$, the model is, again, limited by $\Ra = 2 \times 10^5$.
For high $\Ra$ and $\Pr = 0.025$, the validity boundaries are the BL crossover at low $\Ha \lesssim 10^3$ and the $\Re = 10^6$ boundary for higher $\Ha$.
At $\Pr = 0.005$, the BL crossover is shifted to very high $\Ra > 10^{13}$ (see also Fig.~\ref{fig:phasediagram_mGL}(a)) and the $\Re = 10^6$ boundary is the limiting restriction for increasing $\Ra$ at all $\Ha$.
Since the validity of the quasistatic approximation is dependent on the magnetic Prandtl number, these limits may shift for increasing or decreasing $\Pm$ (the boundary is shifted to smaller or higher $\Ra$, respectively).

The DR contribution crossovers for the kinetic and thermal DR are also plotted in Figs.~\ref{fig:phasediagram_mGL}(b) and~(c) (dash-dotted and dash-double-dotted lines, respectively).
$\varepsilon_{\nu,\mathrm{Bulk}}$ is dominant at low $\Ha$ but is eventually surpassed by $\varepsilon_{\nu,\mathrm{BL}}$ when $\Ha$ increases.
The thermal DR is generally more dependent on $\Ra$ with $\varepsilon_{\kappa,\mathrm{BL}}$ and $\varepsilon_{\kappa,\mathrm{Bulk}}$ being dominant at low and high $\Ra$, respectively.
With decreasing $\Pr$, the kinetc DR crossovers is shifted to higher~$\Ha$. 
The thermal DR crossover at low $\Ha$ shifts to higher~$\Ra$ for decreasing~$\Pr$.
At high $\Ha$, however, it is unaffected by changes in~$\Pr$ and runs parallel to the Chandrasekhar limit.
Since the parameter~$c_4$ is extremely small, the magnetic DR is dominated by $\varepsilon_{\eta,\mathrm{Bulk}}$ for all~$\Ra$ and~$\Ha$.

In summary, the comparison with experimental data (Figs.~\ref{fig:cmp_Nu} to~\ref{fig:cmp_Re}) and the phase diagrams (Fig.~\ref{fig:phasediagram_mGL}) show that despite the reduced complexity of the revised model, its predictions are more accurate and more physically sensible compared to the previous model in Ref.~\onlinecite{Zurner2016c}.

\vspace{-.5ex}  
\section{Conclusion}
\label{sec:conclusion}

\vspace{-1ex}  
An updated model of the heat and momentum transport for Rayleigh-B\'enard convection in a vertical magnetic field was presented.
By revising some of the basic assumptions of the model and including new aspects, the theoretical predictions could be improved significantly.
The inclusion of a generalized kinetic boundary layer thickness allowed for the extension of the model to the low-$\Ha$ limit and to match it with the well established Grossmann-Lohse theory at $\Ha = 0$.
This reduced the complexity of the model greatly by fixing the values of five parameters.
With the removal of the turbulent-to-laminar transition, the total number of free model parameters has thus been reduced from the initial seven to just two.
An extended experimental database also allowed for a more robust fit of the model which effectively removed one more parameter ($c_4 \approx 0$).
Physically, this suggests that the effect of Joule dissipation in the kinetic boundary layer is negligible for the considered parameter range.
The transitions of the boundary layer crossover and the high Prandtl number limit from the Grossmann-Lohse theory are excluded in the present low-$\Pr$ regime.
If the model were to be extended to higher Prandtl numbers, these transitions could be easily reintroduced.
In this case it might also be beneficial to consider a modification of the Grossmann-Lohse ansatz presented by \citet{Bhattacharya2020} (for $\Ha = 0$), who let the model parameters $c_i$ be functions of $\Ra$ and $\Pr$.
This change can produce more accurate predictions especially of the Reynolds number over a wider range of $\Pr$.

The revised model equations satisfactorily reproduce experimental heat transport data for liquid metals at $\Pr\sim0.025$.
Additional experimental data at different Prandtl numbers, for example in liquid sodium with $\Pr \sim 0.005$, would be desirable to further verification of the model.
The momentum transport predictions agree less well with experimental results.
This shows that the suppression of turbulence by the magnetic field cannot be fully reproduced by the Grossmann-Lohse ansatz.
Together with a more rigorous treatment of the weakly non-linear and laminar regimes this is a major challenge for this mean-field theory and should be considered in future investigations.

\begin{acknowledgments}
The author would like to thank J\"org Schumacher for insightful discussions and helpful suggestions.
This work was supported by the Deutsche Forschungsgemeinschaft with grant no.\ GRK~1567.
\end{acknowledgments}

\appendix

\section{The model equations}
\label{apx:modelEq}

This appendix discusses some details of the model equations that were skipped in the main text for the sake of brevity.

\subsection{The initial dissipation rate estimates}
\label{apx:modelEq_DRest}

The scaling relations of the dissipation rate (DR) contributions in~\eqref{eq:DR_estimate} are based on the following arguments, which were introduced in Ref.~\onlinecite{Grossmann2000,Grossmann2001,Zurner2016c}.

In the kinetic boundary layer~(BL), the velocity gradient is estimated as $U/\delta_v$, using the velocity~$U$ of the turbulent large scale wind and the kinetic BL thickness $\delta_v$ as characteristic scales.
With the definition of the viscous DR~\eqref{eq:DR_visc}, the BL contribution amounts to $\varepsilon_{v,\mathrm{BL}} \sim \nu (U/\delta_v)^2$. 
An additional factor is introduced as the volume fraction of the top and bottom BL compared to the whole fluid volume, which amounts to $2\delta_v / H$ (the constant factor of 2 is dropped in the scaling relation).
With an Hartmann layer for the kinetic BL $\delta_v=\delta_{v,B}$, these considerations amount to~\eqref{eq:DR_visc_BL_estimate}.
The bulk flow is assumed to be dominated by turbulence and that the dissipation term $\nu\nabla^2\vec v$ is balanced by the inertial term $(\nabla\cdot\vec v)\vec v$ through an energy cascade.
Using the layer height~$H$ and wind velocity~$U$ as scales, this gives $\nu U/H^2 \sim U^2/H$ and with the definition~\eqref{eq:DR_visc} results in~\eqref{eq:DR_visc_bulk_estimate}.
The bulk volume fraction is approximated as $(H-2\delta_v)/H \sim 1$ for thin BL $\delta_v \ll H$.

The estimates of the magnetic DR~(\ref{eq:DR_magn_bulk_estimate},\ref{eq:DR_magn_BL_estimate}) are calculated using its definition~\eqref{eq:DR_magn}.
As mentioned in point (iii) after equation~\eqref{eq:DR_estimate}, the induced magnetic field strength is estimated as $b\sim\Rm B_0$ and the length scales are the layer height~$H$ and the viscous BL thickness~$\delta_v$ in bulk and BL, respectively.
The volume fractions are applied as before for the viscous DR.

The thermal BL contribution follows from the definition~\eqref{eq:DR_therm}, the linearized temperature gradient $\sim \Delta T/\delta_T$ and the BL volume fraction $\delta_T/H$: $\varepsilon_{\kappa,\mathrm{BL}} \sim \kappa (\Delta T/\delta_T)^2 \delta_T/H = \kappa (\Delta T/H)^2 \Nu$.
If this estimate was used, the first term $\kappa (\Delta T/H)^2$ in \eqref{eq:DR_split_therm} would not be required.
However, \citet{Grossmann2001} further modified this estimate by balancing the advective term $\vec v\cdot\nabla T$ of the heat transfer equation in the bulk and the dissipative term $\kappa \nabla^2 T$ in the BL: $U/H \sim \kappa/\delta_T^2$ which gives $\Nu\sim\sqrt{\Re\Pr}$ and results in the BL contribution estimate~\eqref{eq:DR_therm_BL_estimate}.
The bulk thermal DR~\eqref{eq:DR_therm_bulk_estimate} is determined in direct analogue to the bulk viscous DR.
The bulk volume fraction is again estimated as $(H-2\delta_T)/H \sim 1$ with the thermal BL thickness~$\delta_T$.

It has to be noted that the approximation of the thermal bulk volume fraction $(H-2\delta_T)/H \sim 1$ is not valid close to the onset of convection, where $\delta_T=H/(2\Nu) \to H/2$ since $\Nu\to1$.
However, with an imposed onset of convection (see black lines Figs.~\ref{fig:cmp_Nu_onset} and~\ref{fig:cmp_Re}(b)), this region of the phase space is dominated by the transition function~$h$ and the bulk volume fraction has little influence on the result.
Without an imposed onset of convection (gray lines), the model fails to correctly reproduce the weakly non-linear and laminar regimes which is not changed by including the bulk volume fraction in the model equations.
Additionally, explicit inclusion of the bulk volume fraction would make equation~\eqref{eq:model2} incompatible with the corresponding GL model equation\citep{Stevens2013} and the direct association of parameters~\eqref{eq:param_applyGL} between the models for $\Ha=0$ would be impeded.
As a consequence, the estimate~\eqref{eq:DR_therm_bulk_estimate} is left unchanged.

\subsection{Regime transitions of the initial model}
\label{apx:modelEq_transition}

The transition function $f(x) = (1+x^4)^{-1/4}$ (introduced in Ref.~\onlinecite{Grossmann2001}) has the properties $f(x\to0) \to 1$ and $f(x\to\infty) \to 1/x$.
The term $U f(\delta_{v,B}/\delta_T) = U f(2\Nu/\Ha)$ consequently describes the velocity scale in the thermal boundary layer (BL):
If the viscous BL is nested within the thermal BL ($\delta_T > \delta_{v,B}$) the outer layers of the thermal BL experience the large-scale wind~$U$ and $U f(\delta_{v,B}/\delta_T \to 0) \approx U$.
Conversely, for $\delta_T < \delta_{v,B}$ the velocity scale in the thermal BL is reduced.
Assuming a linear velocity profile in the kinetic BL, this gives a characteristic velocity of $U f(\delta_{v,B}/\delta_T \to \infty) \approx U \delta_T/\delta_{v,B}$.
The estimates~\eqref{eq:DR_therm_bulk_estimate} and~\eqref{eq:DR_therm_BL_estimate} for the initial model thus become $\varepsilon_{\kappa,\mathrm{Bulk}} \sim \kappa (\Delta T/H)^2 \Re\Pr f(2\Nu/\Ha)$ and $\varepsilon_{\kappa,\mathrm{BL}} \sim \kappa (\Delta T/H)^2 \sqrt{\Re\Pr f(2\Nu/\Ha)}$.

Similarly, $g(x) = f(1/x)^{-1}$ has the properties $g(x\to0) \to 1/x$ and $g(x\to\infty) \to 1$.
In the initial model\citep{Zurner2016c}, this allowed $\varepsilon_{v,\mathrm{Bulk}}$ to transition between a turbulent $\Re^3$-scaling ($\Re \gg \Re^\ast$) and a laminar $\Re^2$-scaling ($\Re \ll \Re^\ast$) at a model parameter $\Re^\ast$: $\varepsilon_{v,\mathrm{Bulk}} \sim (\nu^3 / H^4) \Re^3 g(\Re/\Re^\ast)$.

\begin{figure*}
\centering
\includegraphics{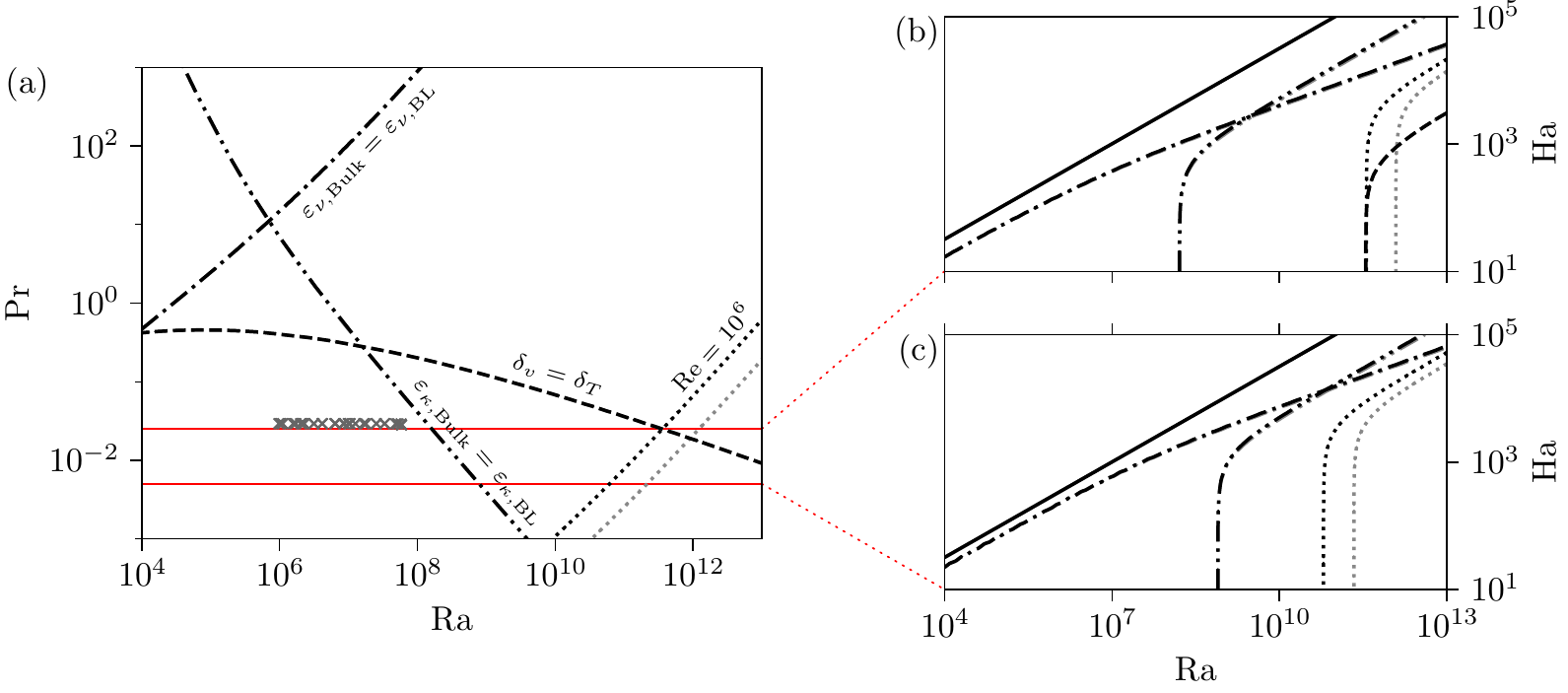}
\caption{%
  Phase diagrams of the model with onset transition and re-scaled parameters~\eqref{eq:param_rescaled}.
  Regime boundaries are plotted as black lines with the same line styles as in Fig.~\ref{fig:phasediagram_mGL}.
  The corresponding regime boundaries with the un-scaled parameters~\eqref{eq:param_applyGL} and~\eqref{eq:param_fit_onset} are replotted from Fig.~\ref{fig:phasediagram_mGL} as gray lines.
  (a)~$(\Ra, \Pr)$ phase diagram at $\Ha = 0$.
  (b)~and (c)~$(\Ra, \Ha)$ phase diagram at $\Pr = 0.025$ and $\Pr = 0.005$, respectively.
  The gray crosses in~(a) mark the positions of the $\Re_\mathrm{LSC}$ measurements\citep{Zurner2019} used to determine $\beta$ with \eqref{eq:Re_rescaleFactor}.}
\label{fig:phasediagram_mGL_rescaled}
\end{figure*}

The third transition function $h(x) = 1-f(x)$ has the properties $h(x\to0) \to 0$ and $h(x\to\infty) \to 1-1/x \to 1$ and is used to enforce the onset of convection\citep{Zurner2016c}.
It is expected that $\Nu \to 1$ below the critical Rayleigh number $\Ra < \Ra_\mathrm{c}$, i.e., in the purely conductive regime.
Let us assume a solution $\Nu$ to a model equation has been found that does not reproduce this behavior (this corresponds to \eqref{eq:model_initial_Nu} without $h$). 
The onset can be imposed by calculating a new result~$\Nu'$ with $\Nu'-1 = (\Nu-1) h(\Ra/\Ra_\mathrm{c})$.
To obtain $\Nu'$ directly from the model, this equation can be inserted into the model equation.
Specifically, every occurrence of $\Nu-1$ is replaced by $(\Nu'-1) / h(\Ra/\Ra_\mathrm{c})$ (the prime of $\Nu'$ is subsequently dropped).
The argument of the transition function $f(2\Nu/\Ha)$ is not modified, since $h$ is only relevant for $\Ra$ close to or below $\Ra_\mathrm{c}$ in which case $\Nu$ is small, $\Ha$ is large and $f(2\Nu/\Ha) \approx 1 = \mathrm{const}$.
$h$ is not introduced into the model equation for $\Re$, since $\Re\to0$ follows intrinsically from~\eqref{eq:model_initial_Re} for $\Nu \to 1$.

\section{Reynolds number re-scaling}
\label{apx:Re_rescale}

The model equations~\eqref{eq:model1} and~\eqref{eq:model2} are invariant under the transformations
\begin{align}
\label{eq:param_transform}
\begin{aligned}
\Re &\to \beta \Re \,, & 
a &\to \beta^{1/2} a \,, \\
c_1 &\to \beta^{-3} c_1 \,, & \quad
c_2 &\to \beta^{-2} c_2 \,, & \quad
c_3 &\to \beta^{-2} c_3 \,, \\
c_4 &\to \beta^{-2} c_4 \,, &
c_5 &\to \beta^{-1} c_5 \,, &
c_6 &\to \beta^{-1/2} c_6
\end{aligned}
\end{align}
for any $\beta\in\mathbb{R}$.
This means that the Reynolds number can be re-scaled by an arbitrary factor without affecting the result of the Nusselt number.
Fitting the model equations~\eqref{eq:model1} and~\eqref{eq:model2} to heat transfer data $(\Ra, \Ha, \Pr, \Nu)$ can thus result in an infinite number of fit values $a$ and $c_1$ to $c_6$ that are consistent with the transformations~\eqref{eq:param_transform}, give exactly the same result for $\Nu$, but result in wildly different Reynolds numbers for a given point $(\Ra, \Ha, \Pr)$.
Note, that the shape of the function $\Re(\Ra, \Ha, \Pr)$ over the phase-space is the same for all these fit results and is just shifted by a constant factor.

In Ref.~\onlinecite{Zurner2019} it was shown, that the Reynolds number of the GL theory underpredicts the experimental Reynolds number~$\Re_\mathrm{LSC}$ based on the velocity of the large-scale circulation~(LSC), i.e., the convective wind.
To adapt $\Re$ of the model to the experimental data, the factor~$\beta$ is determined by
\begin{align}
\label{eq:Re_rescaleFactor}
\beta &= \frac{\Re_\mathrm{LSC}}{\Re(\Ra, \Ha, \Pr)} \,,
\end{align}
where $\Re$ is calculated with model parameters~\eqref{eq:param_applyGL} and~\eqref{eq:param_fit_onset} at the point $(\Ra, \Ha, \Pr)$ at which $\Re_\mathrm{LSC}$ was measured ($\Ha = 0$ and $\Pr = 0.029$ for all measurements in Ref.~\onlinecite{Zurner2019}).
The average result is $\beta = 1.81 \pm 0.08$ and the resulting re-scaled model parameters are
\begin{equation}
\label{eq:param_rescaled}
\begin{aligned}
c_1 &= 0.233 \,, &
c_2 &= 2.26 \,, &
c_3 &= 0.0137 \,, \\
c_4 &= 2.30 \times 10^{-18} \,, &
c_5 &= 0.0139 \,, &
c_6 &= 0.362 \,, \\
&& a &= 1.24 \vphantom{10^{-18}} \,.
\end{aligned}
\end{equation}
Propagation of the standard deviations from $\beta$, $c_3$ and $c_4$ results in uncertainties of 0.031 for $c_1$, 0.20 for $c_2$, 0.0018 for $c_3$, $8.36\times 10^{-7}$ for $c_4$, 0.0006 for $c_5$, 0.008 for $c_6$ and 0.03 for $a$.

While the Nusselt number is unchanged, this scaling of $\Re$ has to be considered for the regime boundaries.
Figure~\ref{fig:phasediagram_mGL_rescaled} shows the regime boundaries of the re-scaled model (black lines) in comparison to the un-scaled model (gray lines, re-plot of the black lines in Fig.~\ref{fig:phasediagram_mGL}).
The BL crossover and DR contribution crossovers are unchanged, as they are invariant under the transformations~\eqref{eq:param_transform} as well.
However, the $\Re = 10^6$ boundary is shifted to smaller $\Ra$.
As a result, for $\Pr = 0.025$ (Fig.~\ref{fig:phasediagram_mGL_rescaled}(b)) the BL crossover and $\Re = 10^6$ coincidentally take place at the same $\Ra$ for $\Ha \lesssim 10^2$.

The Reynolds number are adjusted by a constant factor only, since the GL ansatz assumes that all dependencies on the control parameters are covered by the base equations~\eqref{eq:base_relations} and the dissipation rate estimates including transition functions. 
Thus, the re-scaled parameters~\eqref{eq:param_rescaled} are only valid for small Prandtl numbers.
To get correct results for low-, intermediate- and high-$\Pr$ regimes, the GL theory would need to be revisited and revised to properly reproduce all these cases.
A possible solution for this issue is presented by \citet{Bhattacharya2020}, who modified the GL approach by introducing non-constant parameters $c_i(\Ra,\Pr)$ for the case $\Ha = 0$.
This alternative approach produced Reynolds number predictions that coincided better with numerical and experimental data over a large range of~$\Pr$.

\bibliography{Manuscript_Bibliography}

\end{document}